\documentclass[12pt,a4paper,twoside]{article}
\RequirePackage{latexsym,amsmath,amssymb}
\RequirePackage[dvipsnames,usenames]{color}
\usepackage{a4wide}
\usepackage{amssymb}
\usepackage{bm}
\usepackage{dcolumn}
\usepackage{amsfonts}
\usepackage{graphicx,epsfig}
\usepackage{psfrag}
\usepackage{cite}
\usepackage{multicol}
\usepackage{tikz}
\usetikzlibrary{arrows,shapes}
\usetikzlibrary{matrix,arrows}
\newcommand{\be}{\begin{eqnarray}}
\newcommand{\ee}{\end{eqnarray}}

\renewcommand{\d}{\mbox{${\rm d}$}} 
\newcommand{\lp}{\ell_{\rm p}}
\newcommand{\mpl}{m_{\rm p}}
\newcommand{\gn}{G_{\rm N}}

\title{\bf Bootstrapped Newtonian cosmology and the cosmological constant problem}
\author{Roberto~Casadio$^{ab}$\thanks{E-mail: casadio@bo.infn.it}
$\,$and
Andrea~Giusti$^{c}$\thanks{E-mail: agiusti@ubishops.ca}
\\
\\
$^a${\em Dipartimento di Fisica e Astronomia, Universit\`a di Bologna}
\\
{\em via Irnerio~46, I-40126 Bologna, Italy}
\\
\\
$^b${\em I.N.F.N., Sezione di Bologna, IS - FLAG}
\\
{\em via B.~Pichat~6/2, I-40127 Bologna, Italy}
\\
\\
$^c${\em Department of Physics and Astronomy, Bishop's University,}
\\
{\em 2600 College Street, Sherbrooke, Qu\'ebec, Canada J1M 1Z7}
}
\begin{document}
\maketitle
%
\begin{abstract}
Bootstrapped Newtonian gravity was developed with  the purpose of estimating the impact
of quantum physics in the nonlinear regime of the gravitational interaction, akin to corpuscular models
of black holes and inflation.
In this work, we set the ground for extending the bootstrapped Newtonian picture to cosmological spaces.
We further discuss how such models of quantum cosmology can lead to a natural solution of the cosmological
constant problem.
\end{abstract}
\newpage
%
%
%
%
%
%
%
\section{Introduction}
\label{intro}
\setcounter{equation}{0}
The standard model of the present Universe, often referred to as the $\Lambda$-Cold Dark Matter
($\Lambda$CDM) model, allows us to include the current accelerated expansion of the Universe,
confirmed by type I supernovae surveys~\cite{Perlmutter:1998np}, in the broader scheme of General Relativity.
However, this is carried out at the price of adding an {\em ad hoc} cosmological constant term
$\Lambda$ in the Einstein field equations or by assuming the existence of an exotic fluid, dubbed
{\em dark energy}, whose pressure $p$ and energy density $\rho$ are such that
$\rho + p \approx 0$.~\footnote{Throughout the work we set the speed of light $c = 1$, while we express
Newton's constant $\gn=\lp/\mpl$ and Planck's constant $\hbar=\lp\,\mpl$ in terms of the Planck length $\lp$
and the Planck mass $\mpl$. Further, we use $\simeq$ to denote that two quantity are equal
up to order-one multiplicative factors.} 
\par
Corpuscular gravity~\cite{dvali-1, dvali-2, giusti-rev} is a quantum description based on the
{\em classicalization} scheme~\cite{classicalization} applied to the gravitational interaction.
In some detail, the ultraviolet incompleteness of the Einstein-Hilbert gravity is conjectured to be cured
by the formation of extended bound states of a large number of soft (off-shell) gravitons in very energetic
processes.~\footnote{For interesting cases with a few gravitons involved see Refs.~\cite{blas}}
These gravitons are prone to escape the bound state by quantum depletion, since they
are very weakly coupled with each other and to other fields,
but the large occupation endows the collective quantum state with a quasi-classical behaviour.
According to this paradigm, a black hole of mass $M$ is understood as a marginally bound state of
$N \simeq M^2 / \mpl^2$ off-shell gravitons of typical momentum $\epsilon _{\rm G} \simeq \mpl / \sqrt{N}$
interacting with each other with an effective coupling $\alpha _{\rm G} \simeq 1/N$.
Since the collective coupling $N \, \alpha _{\rm G} \simeq 1$, the bound state cannot hold on to its
constituents and will slowly lose components by the process of quantum depletion.
The depletion of constituent gravitons represents the quantum precursor of the semiclassical
Hawking evaporation as a consequence of the softness of the constituents and combinatorics.
In fact, the master equation of the Hawking emission is recovered in the semiclassical limit
of the theory, {\em i.e.}~for $N \to \infty$ and $\lp \to 0$, with the length scale
\be
L
=
\sqrt{N} \, \lp
\label{eq:LN}
\ee
and $\hbar$ kept fixed and finite.
In the language of Bose-Einstein condensates, a ``corpuscular'' black hole is a condensate of gravitons
on the verge of a quantum phase transition, which remains at the critical point throughout the evaporation
process, up until the quantum break-time of the system.
This ``quantum criticality''  provides nearly-gapless  Bogoliubov modes (with energy gap $\Delta \epsilon \sim 1/N$)
accounting for the entropy of the black hole.
A similar scenario can be derived from the Hamiltonian approach to gravity for the gravitational
collapse of compact spherically symmetric distributions of matter~\cite{baryons}.
Further, it is worth noting that this peculiar quantum description of black holes is strictly intertwined with
the generalised uncertainty principle~\cite{Luca-1} and it has interesting implications for black-hole
echoes~\cite{Luca-2}.  
\par
Other space-times equipped with horizons have been studied within this scenario.
In particular, one can apply arguments similar to the one above for black holes to the de~Sitter
space~\cite{dvali-2, Casadio:2017twg, sardine-1, sardine-2, sardine-3, sardine-4},
embedding inflation and the late-time evolution of the Universe within a single theoretical paradigm.
This corpuscular picture of the Universe further allows one to explain the dark matter phenomenology
as the response of the cosmological condensate of gravitons, responsible for the quasi-de~Sitter
behaviour at large scale, to the local presence of impurities, {\em i.e.}~galaxies and (possibly)
clusters of galaxies.
Additionally, the corpuscular Hamiltonian approach~\cite{baryons} for cosmological spaces
provides a quantum exclusion principle for the de~Sitter space~\cite{dvali-2, Casadio:2017twg}
and phantom energy~\cite{corpuscularValerio}.
\par
In a nutshell, bootstrapped Newtonian gravity~\cite{PN, boot-1, boot-2, boot-3, boot-4} is an approach
aimed at providing a more detailed field-theoretic realisation for the quantum corpuscular theory of
(spherically symmetric) compact sources.
This approach has at its core the idea that, if quantum effects were negligible, starting from a field theory
reproducing Newton's gravity, one should be able to recover General Relativistic configurations by
reconstructing (leading-order) non-linearities out of the coupling of each constituent graviton with
the collective state.
However, quantum effects are not assumed to be negligible and departures from General 
Relativity are therefore expected (see, {\em e.g.}~Refs.~\cite{giusti-rev, PN, boot-1} for more details).
This procedure represents, to some extent, an analogue of Deser's conjecture that the Einstein-Hilbert
action in the vacuum should be obtained from a Fierz-Pauli Lagrangian for spin 2 fields in Minksowski space-time
by adding suitable self-coupling terms~\cite{deser-1, deser-2}.
Our approach instead focuses on the interplay between non-linearities and quantum physics in order to 
determine possible quantum corrections to the predictions of General Relativity for compact but
large self-gravitating systems in the strong field regime, for which the matter content is
also far from the quantum vacuum.
\par
In this work, we lay the basis for an extension of the bootstrapped Newtonian picture to cosmology,
where the much simpler homogenous and isotropic spaces allow us to analyse also higher order terms.
Furthermore, we show how this novel paradigm provides a natural solution to the cosmological
constant problem.
\section{Corpuscular models}
\label{sez:cc}
\setcounter{equation}{0}
Let us start from the assumption that matter and the corpuscular state of gravitons together
must reproduce the Einstein equations.
In particular, we will focus on the Hamiltonian constraint corresponding
to the time-reparameterisation invariance of the theory~\footnote{The Hamiltonian constraint 
also plays the role of ensuring total energy conservation in a suitable sense.}
\be
H_{\rm M}+H_{\rm G}
=
0
\ ,
\label{H0}
\ee
where $H_{\rm M}$ is the matter energy and $H_{\rm G}$ the analogue quantity
for the graviton state.
We recall that local gravity being attractive in general implies that $H_{\rm G}\le 0$,
although this is not true for the graviton self-interaction~\cite{baryons,PN}.
\subsection{Corpuscular De~Sitter and black holes}
The corpuscular models of gravity of Refs.~\cite{dvali-1, dvali-2, giusti-rev} were first introduced for
describing black holes but are perhaps easier to obtain in cosmology, as the two systems share a
fundamental ``critical'' condition.
In order to have the de~Sitter space-time in General Relativity, one must assume the existence of a
cosmological constant term, or vacuum energy density $\rho$, so that the Friedman equation 
(for spatially flat Universe) reads
\be
3\left(\frac{\dot a}{a}\right)^2
=
8\,\pi\,\gn\,\rho
\ .
\label{eq:FL}
\ee
Upon integrating on the volume inside the Hubble radius 
\be
L
=
\frac{a}{\dot a}
\ ,
\label{Laa}
\ee
we obtain
\be
L
\simeq
\gn\,L^3\,\rho
\simeq
\gn\,M
\ .
\label{LgM}
\ee
This relation looks like the expression of the horizon radius for a black hole of ADM mass
$M$ and is the reason it was conjectured that the de~Sitter space-time could likewise
be described as a condensate of gravitons~\cite{dvali-2}.
\par
One can describe the ideal corpuscular model as a condensate of $N$ (soft off-shell) gravitons of typical
Compton length equal to $L$ in Eq.~\eqref{eq:LN}.
Then, since $M = N \, \epsilon _{\rm G}$ and $\epsilon _{\rm G}= \hbar / L$, one can infer that
\be
M
=
N\,\frac{\lp\,\mpl}{L}
=
\sqrt{N}\,\mpl
\ ,
\label{HL}
\ee
in agreement with Eq.~\eqref{LgM}.
This shows that for a macroscopic black hole, or de~Sitter universe, one needs a huge number $N\gg 1$.
Since the above corpuscular relations are more realistically expected to hold only approximately,
we will assume that the energy $M$ is exactly given in terms of the fixed number of gravitons
$N$ by Eq.~\eqref{HL}, whereas the Compton length of the gravitons denoted by $L$
henceforth can be (slightly) different from the Hubble value in Eq.~\eqref{eq:LN},
which we will denote with $L_N=\gn\,M=\sqrt{N}\,\lp$. 
We thus write the Compton length as
\be
L
=
\xi^{-1}\,{\sqrt{N}\,\lp}
\equiv
\xi^{-1}\,{L_N}
\ ,
\label{LMn}
\ee
where $\xi$ is a positive constant of order one which will play a significant role in the following
analysis.
In particular, we anticipate that we will find $L\gtrsim L_N$, that is $\xi\lesssim 1$, so that the
gravitons must spread outside the Hubble radius $L_{\rm N}$.
\subsection{``Post-Newtonian'' corpuscular models}
Let us refine the above corpuscular description.
In Refs.~\cite{baryons,PN}, it was shown that the maximal packing condition which yields
the scaling relations~\eqref{LMn} for a black hole actually follow from the energy
balance~\eqref{H0} when matter becomes totally negligible.
If $H_{\rm M}=0$, one is left with
\be
\label{eq:Hconstraint}
H_{\rm G}
=
U_{\rm N}
+
U_{\rm PN}
=
0
\ ,
\label{H00}
\ee
where $H_{\rm G}$ is given by the sum of the negative~\footnote{Of course, we 
must assume the gravitational interaction is attractive and it is then important to keep track
of minus signs in the following expressions.} Newtonian energy~\cite{baryons}
\be
U_{\rm N}
\simeq
N\,\epsilon_{\rm G}
\simeq
-N\,\frac{\lp\,\mpl}{L}
\simeq
-\frac{\lp\,M^2}{L\,\mpl}
\ ,
\label{UN0}
\ee
and the positive ``post-Newtonian'' contribution~\cite{baryons}
\be
U_{\rm PN}
\simeq
N\,\frac{\sqrt{N}\,\lp^2\,\mpl}{L^2}
\simeq
\frac{\lp^2\,M^3}{L^2\,\mpl^2}
\ .
\label{UPN0}
\ee
One therefore immediately recovers the scaling relation~\eqref{LMn} from Eq.~\eqref{H00},
regardless of whether $L$ is the horizon radius of the de~Sitter universe or of a black hole.
\section{Bootstrapped Newtonian series}
\label{sez:higher}
\setcounter{equation}{0}
Let us now note that 
\be
U_{\rm N}
\simeq
\left(
-\frac{\sqrt{N}\,\lp}{L}
\right)
M
=
-\xi \,M
\ ,
\ee
where $\xi$ was introduced in Eq.~\eqref{LMn}, and
\be
U_{\rm PN}
\simeq
-\xi\,U_{\rm N}
\simeq
\frac{L_N\,M}{L^2}
\ ,
\ee
where we remark that $L$ is the Compton length of gravitons associated with
either the horizon radius $L_N$ of the de~Sitter universe or of a black hole.
This leads us to conjecture that the complete form of the gravitational Hamiltonian
to all orders in the $\xi$ expansion is given by
\be
H_{\rm G}
\simeq
M\,
\sum_{n = 1}^\infty
g_{n} 
\left(-\xi\right)^{n}
\ ,
\label{exp}
\ee
where we notice that $g_1\simeq g_2\simeq 1$ and
the alternating signs above are precisely due to the attractiveness of gravity.
This sort of expansions is strictly related to the argument of resurgence in QFT.
In particular, the values of $g_{n}$ should arise directly from this argument
(see, {\em e.g.}~Ref.~\cite{resurgence}).
\par
For a black hole, a matter contribution is needed and we cannot have
$H_{\rm G}=0$.
In particular, assuming the scaling relation~\eqref{LMn}, we note in passing that
Eq.~\eqref{exp} would reproduce the Newtonian expectation,~\footnote{We recall
that the horizon radius can in fact be interpreted in terms of Newtonian gravity
as the size of a source with escape velocity  equal to the speed of light.}
that is
\be
H_{\rm G}
\simeq
U_{\rm N}
\ ,
\ee
for $g_n\simeq e\,\xi^{1-n}/n! (e-1)$.~\footnote{Note that in this
``Newtonian'' case, the condition $g_1\simeq g_2\simeq 1$ does not hold.}
Eq.~\eqref{H0} then yields $H_{\rm M}\simeq U_{\rm N}$
with this expression of the coefficients $g_n$.
\par
One more naive observation is that the typical momentum of the constituent gravitons scales as $1/L$
which, in turn, is related to the spectral representation of the differential operator ${\d}/{\d r}$.
Hence, one could view the exponential formula~\eqref{exp} as the realisation of the black hole quantum
state in infinite derivative gravity~\cite{yung}.
More precisely, one can see that the Schwarzschild metric (or quantum modifications thereof,
and perhaps quantum General Relativity as a whole), should be recovered as an infinite derivative theory
in flat space-time (where the corpuscular picture is naturally formulated).
Again, this is in line with bootstrapped Newtonian gravity~\cite{PN, boot-1} and Deser's
works~\cite{deser-1, deser-2}.
\par
From now on we shall focus on pure gravity and cosmology, for which Eq.~\eqref{exp} is more suitable. 
Let us then try to keep the analysis of Eq.~\eqref{exp} as general as possible.
We only require $g_n>0$ in the series~\eqref{exp} and the ``on-shell'' condition
\be
\label{stocazzo}
H_{\rm G}(L)
=
0
\quad
{\rm for}
\quad
L
=
\bar L
=
\xi^{-1}\,L_N
\ ,
\ee
with $\xi>0$ (and of order one) in the scaling relation~\eqref{LMn}.
First of all, the series~\eqref{exp} converges provided the coefficients $g_n$ satisfy the inequality
\be
\label{eq:17}
g_{n+1}
<
\xi^{-1}\,g_n
\ .
\ee
Note that, for our purpose, convergence is strictly required only for $L=\bar L$. 
Assuming however the series converges for $L$ at least in a neighbourhood of $\bar L$,
the Hamiltonian $H_{\rm G}$ must be a smooth function of $L$ around $\bar L$,
that is
\be
\left.
\frac{\partial^k H_{\rm G}}{\partial L^k}
\right|_{L=\bar L}
=
\left.
M\,
\sum_{n = 1} ^\infty g_{n}\, 
\frac{\partial^k}{\partial L^k}
\left(-\frac{L_N}{L}\right)^n
\right|_{L=\bar L}
\ ,
\ee
for some integers $1\le k \le K$.
Approaching the cosmological (or black hole) horizon,
the contributions of nonlinear order of the theory will become of the same order of the
Newtonian one, signalling the beginning of the full non-perturbative regime of gravity.
This suggests that, not only $g_1\simeq g_2\simeq 1$ but also $g_{n} = \mathcal{O}(1)$
for all $n\geq 1$ (when $L$ approaches $L_{N}$), which is indeed consistent with Eq.~\eqref{eq:Hconstraint}.
Putting together this observation with the condition in Eq. \eqref{eq:17}, the convergence of
the above series requires $\xi$ to be smaller of an $\mathcal{O}(1)$ constant which depends
on the specific behaviour of the sequence of coefficients $g_n$.
We remark that we always assume that the energy $M$ is fixed by the number of gravitons
according to Eq.~\eqref{HL} and it therefore does not depend on the Compton length $L$
(again for values around $L_N$).
We then expect to obtain (slightly) more restrictive conditions for the coefficients $g_n$.
\par
At this point, it is more interesting to consider other quantities of physical interest, and employ their
relations with $H_{\rm G}$, in order to obtain further conditions on the coefficients $g_n$.
For instance, we know that the pressure $p=-\rho$ in pure de~Sitter space-time.
From 
\be
p
=
\left.
\frac{1}{4\,\pi\,L^2}
\frac{\partial H_{\rm G}}{\partial L}
\right|_{L=\bar L}
&\!\!=\!\!&
\left.
\frac{M}{4\,\pi\,\bar L^2}\,
\sum_{n = 1} ^\infty g_{n}\, 
\frac{\partial}{\partial L}
\left(-\frac{L_N}{L}\right)^n
\right|_{L=\bar L}
\nonumber
\\
&\!\!=\!\!&
-\frac{M}{4\,\pi\,\bar L^3}\,
\sum_{n = 1} ^\infty 
n\,g_{n}
\left(-\xi\right)^n
\ ,
\label{pressureseries}
\ee 
where we again used Eq.~\eqref{HL}.
On considering the first two terms only ($n=1,2$) and $H_{\rm G}=0$ up to the same order, 
that is $g_1 = \xi\,g_2$, one finds
\be
p
=
-\frac{M\,g_2\,\xi^2}{4\,\pi\,\bar L^3}
=
-\frac{g_2\,\mpl\,\xi^5}{4\,\pi\,\lp^3\,N}
\ .
\ee
From Eqs.~\eqref{LgM}, \eqref{HL} and~\eqref{stocazzo}, we find
\be
\rho
=
\frac{M}{4\,\pi\,\bar L^3}
=
\frac{\mpl\,\xi^3}{4\,\pi\,\lp^3\,N}
\ ,
\label{rhoN}
\ee
so that $p=-\rho$ implies
\be
g_2
=
\xi^{-2}
=
g_1^2
\ .
\ee
Upon replacing this solution into Eq.~\eqref{exp}, we obtain the (slightly) off-shell Hamiltonian
\be
H_{\rm G}(L)
=
M
\left[
-
\xi^{-1}\left(\frac{\sqrt{N}\,\lp}{L}\right)
+
\xi^{-2}
\left(\frac{\sqrt{N}\,\lp}{L}\right)^2
+
\sum_{n = 3} ^\infty
\, g_{n} 
\left(-\frac{\sqrt{N}\,\lp}{L}\right)^n
\right]
\ ,
\label{eq:HGoff}
\ee
where the higher-order terms should progressively capture all the nonlinearities
of the quantum theory of gravity.
By looking at Eq.~\eqref{eq:HGoff}, we can furthermore conjecture that
$g_n\simeq\xi^{-n}$, but the precise values of such coefficients of interest to us will have to
entail the specific features of the actual quantum state of the Universe.
The precise determination of the $g_n$ for the observable Universe hence requires further investigation,
as they will necessarily depend on the actual matter distribution and time evolution which led
to the present state.
\par
In particular, since our Universe is not empty now, we can employ a more general equation
of state for the purpose of improving our estimates of the coefficients $g_n$.
In fact, we can easily generalise the above picture for any equation of state of the form
\be
p
=
-\omega\,\rho
\ ,
\ee
where $0<\omega\le 1$ is a constant.
This yields
\be
\frac{g_2}{\omega}
=
\xi^{-2}
=
\frac{g_1^2}{\omega^2}
\ .
\label{eq:omega}
\ee
We can then conjecture that $g_n\simeq\omega\,\xi^{-n}$.
Note that the {\em pure dust} case $p=0$ turns out to be rather peculiar in this scenario.
Indeed, there are two possible configurations that allow Eq.~\eqref{pressureseries}
to reduce to $p=0$: i) $g_n = 0$, $\forall n \geq 2$, {\em i.e.}~the purely Newtonian case,
in agreement with the Newtonian derivation of the Friedmann equation~\eqref{eq:FL};
ii) $\xi = 0$, which implies $N \to 0$, thus reducing to Minkowski space.
\par
We can finally notice that, upon employing the present value for the Hubble size
of the Universe, that is $L_{N}\simeq \bar L\simeq 10^{27}\,$m, Eq.~\eqref{rhoN}
yields the correct order of magnitude for the present total energy density
\be
\rho
=
\frac{M}{4\,\pi\,\bar L^3}
\simeq
10^{-28}\,\frac{\rm kg}{{\rm m}^3}
\ ,
\ee
without any further assumptions.
From Eq.~\eqref{rhoN} and the definition~\eqref{LMn} we can additionally estimate the
observed value
\be
\xi
=
\left(4\,\pi\,\rho\,\frac{\lp}{\mpl}\,L_N^2\right)^{1/3}
\simeq
0.9 < 1
\ ,
\ee
which implies that the Compton length of the background gravitons $L > L_N$, as anticipated.
Eq.~\eqref{eq:omega} can then be implemented to infer the equation of state
parameter $\omega\simeq \xi\,g_1\simeq 0.9\,g_1$ and $g_2\simeq g_1/\xi\simeq 1.1\,g_1$,
where we again recall that $g_1\simeq 1$.
In fact, all of these final estimates are consistent with our previous assumption that
the coefficients $g_n=O(1)$.
\par
The above (somewhat preliminary) analysis shows that the apparently unnatural small value of the 
cosmological constant is indeed consistent with the quantum state of the observable Universe obtained 
from the bootstrapped Newtonian approach (see also Refs.~\cite{dvali-2,Dvali:2020etd}).
Moreover, a detailed comparison with cosmological measurements can
be employed in order to determine the fine details of the quantum
state of gravitons forming the background of the present quasi-de~Sitter Universe.
Of course, one would need to consider the time evolution of the Universe
from the surface of last scattering in order to determine the particle horizon.
We leave such a question for further developments.
\section{Concluding remarks}
\label{sez:conc}
\setcounter{equation}{0}
Inspired by the corpuscular theory of gravity, in which the (matter empty) de~Sitter Universe is 
described by a bound state of gravitons, we have shown how one can express
the gravitational Hamiltonian $H_{\rm G}$ for a cosmological space to all ``post-Newtonian''
orders as the Newtonian energy of the leading order $U_{\rm N}$ multiplied
by a power series in the quantity $\xi=L_N / L$, where $L$ is the Compton
length of the bound gravitons and $L_N$ is the Hubble radius of the Universe determined
by its (vacuum) energy $M$.
In other words, similarly to Deser's classical argument~\cite{deser-1,deser-2},
one can reconstruct the full quantum Hamiltonian of cosmology by supplying
(``bootstrapping'') the Newtonian potential energy with the energies due to particle
self-interactions within the marginally bound state.
At the fundamental level, this further suggests that General Relativity
might be recovered as an infinite derivative field theory on a flat space-time,
although quantum corrections are naturally expected to introduce deviations
from the classical configurations, which will result in different values of the
coefficients in the ``post-Newtonian'' expansion~\eqref{exp}.
\par
A crucial role in our analysis is played by the parameter $\xi$, which measures the (slight)
departure from the ideal corpuscular relation $L=L_N$.
A functional behaviour for the coefficients of the bootstrapped Newtonian
expansion of $H_{\rm G}$ in powers of $\xi^{-1}$ around the ideal de~Sitter case
with $\omega=1$ was then obtained in Eq.~\eqref{eq:HGoff}.
In order to accommodate for our non-empty observable Universe, we allowed for a more general
equation of state. 
In particular, if one sets the radius $L_N\simeq L$ of the cosmological horizon to the present
size of the Universe, the corpuscular Hamiltonian constraint of cosmology yields
an energy density that matches the observed one associated with the cosmological
constant provided the parameter $\xi\simeq 0.9<1$.
This implies that $L\gtrsim L_N$ and the gravitons spread slightly beyond the measured
Hubble size.
Hence, this prediction for the vacuum energy density coming solely from the
observed size of the Universe and the corpuscular quantum description of
cosmological spaces can arguably be regarded as leading to a natural solution of the
cosmological constant problem.
\section*{Acknowledgments}
R.C.~is partially supported by the INFN grant FLAG.
A.G.~is supported by the Natural Sciences and Engineering Research Council of
Canada (Grant No.~2016-03803 to V.~Faraoni) and by Bishop's University. 
This work has been carried out in the framework of activities of the
Italian National Group for Mathematical Physics [Gruppo Nazionale per la Fisica
Matematica (GNFM), Istituto Nazionale di Alta Matematica (INdAM)]
and COST action {\em Cantata}.
\end{document}